\documentclass[prb,twocolumn,showpacs,amsmath,amssymb,superscriptaddress]{revtex4}
\usepackage{graphicx}
\usepackage{bm}

\begin{document}

\title{Radio--Frequency Method for Investigation of Quantum Properties of Superconducting Structures}
\author{E. Il'ichev}
\email{ilichev@ipht-jena.de}
\affiliation{%
Institute for Physical High Technology, P.O. Box 100239, D-07702
Jena, Germany}
\author{A.Yu.\ Smirnov}
\affiliation{D-Wave Systems Inc., 320-1985 W. Broadway, Vancouver,
B.C., V6J 4Y3, Canada}
\author{M. Grajcar}
\affiliation {Department of Solid State Physics, Comenius
University, SK-84248 Bratislava, Slovakia}
\author{A. Izmalkov}
\affiliation{%
Institute for Physical High Technology, P.O. Box
100239, D-07702 Jena, Germany}
\affiliation {%
Moscow Engineering Physics Institute (State University),
Kashirskoe sh. 31, 115409 Moscow, Russia}

\author{D.~Born}
\affiliation{%
Institute for Physical High Technology, P.O. Box
100239, D-07702 Jena, Germany}
\affiliation{Friedrich Schiller University, Institute of Solid State
Physics, D-07743 Jena, Germany}
\author{N.~Oukhanski}
\affiliation{%
Institute for Physical High Technology, P.O. Box 100239, D-07702
Jena, Germany}
\author{Th.~Wagner}
\affiliation{%
Institute for Physical High Technology, P.O. Box
100239, D-07702 Jena, Germany}
\author{W.~Krech}
\affiliation{Friedrich Schiller University, Institute of Solid State
Physics, D-07743 Jena, Germany}
\author{H.-G.~Meyer}
\affiliation{%
Institute for Physical High Technology, P.O. Box 100239, D-07702
Jena, Germany}
\author{A. Zagoskin}
\affiliation{D-Wave Systems Inc., 320-1985 W. Broadway, Vancouver,
B.C., V6J 4Y3, Canada}
\affiliation{%
Physics and Astronomy Dept., The University of British Columbia,
6224 Agricultural Rd., Vancouver, B.C., V6T 1Z1 Canada}
\date{\today}

\begin{abstract}

We implement the impedance measurement technique (IMT) for
characterization of interferometer-type superconducting qubits. In
the framework of this method, the interferometer loop is inductively
coupled to a high-quality tank circuit. We show that the IMT is a
powerful tool to study a response of externally controlled two-level
system to different types of excitations. Conclusive information about
qubits is obtained from the read-out of the tank properties.

\end{abstract}


\maketitle

\section{Introduction}
Quantum effects in mesoscopic superconducting circuits of
small Josephson junctions have attracted renewed attention. It was
clearly demonstrated that Josephson devices can behave like single
microscopic particles if they are sufficiently isolated from the
environment. Therefore, ideas
developed in atomic and molecular physics can be used for
description of artificially fabricated circuits of macroscopic
size. These concepts are stimulated further by the perspectives of
a promising way to realize quantum bits (qubits) for quantum
information processing.

Qubits are two-level quantum systems with
externally controlled parameters. Generally, two kinds of such
devices with small-size Josephson junctions have been
developed. One approach is based on charge degree of freedom, basis
states of this kind of qubits are distinguished by the number of
Cooper-pairs on a specially designed island. The alternative
realization utilizes the phase of a Josephson junction (or the
flux in a ring geometry), which is conjugate to the
charge degree of freedom.
Due to  macroscopic size of superconducting qubits, they are
extremely sensitive to external disturbances. Thus, a backaction
of a detector should be as small as possible. A lot of different
detectors have been suggested in literature (see
Ref.~\onlinecite{dima} and references therein).

In this paper we review our results obtained on superconducting qubits
by the impedance measurement technique (IMT). Below
we shall discuss several quantum effects including macroscopic
quantum tunneling, Landau-Zener transitions, Rabi oscillations,
and direct resonant spectroscopy of the qubit energy levels. Finally,
we  present our very recent results of investigation of two coupled qubits.

\section{Macroscopic Quantum Tunneling}
For the flux qubits the Josephson energy dominates over the charge
energy, $E_J \gg E_C$. It was predicted, that such systems should
exhibit various quantum-mechanical effects including macroscopic
quantum tunneling (MQT) of the flux.~\cite{legg} Indeed, predicted
effects had been observed
experimentally.~\cite{Friedman00,Wal00,Clarke,Rouse} In this
section we briefly discuss the main properties of the flux qubits
and demonstrate that the IMT technique is a powerful tool for the
investigation of the MQT.

One of the realizations of the flux qubit is a superconducting
loop with low inductance $L_q$, including three Josephson
junctions (a 3JJ qubit).~\cite{Mooij99} Its total Josephson energy
is $E_\mathrm{J} = \sum_{i=1}^3E_{\mathrm{J}i}(\phi_i)$, where
$\phi_i$ and $E_{\mathrm{J}i}=\hbar I_{\mathrm{c}i}/2e$ are the
phase difference and Josephson energy of the $i$th junction,
respectively. Due to flux quantization, only $\phi_{1,2}$ are
independent with
$\phi_3=-\phi_1-\phi_2-2\pi\Phi_\mathrm{e}/\Phi_0$ ($\Phi_0=h/2e$
is the flux quantum) for  negligibly small~$L_q$ (though see
\cite{Alec03}).

At $\Phi_\mathrm{e}=\frac{1}{2}\Phi_0$, the potential $U(\phi_1,\phi_2)$ has
two shallow minima. These two minima correspond to the qubit states
$\psi_L$ and $\psi_R$, carrying opposite but equal supercurrents around the
loop. Therefore, according to the laws of quantum mechanics,
near degeneracy the system can tunnel between the two potential minima.

\begin{figure}
\includegraphics[width=8.1cm]{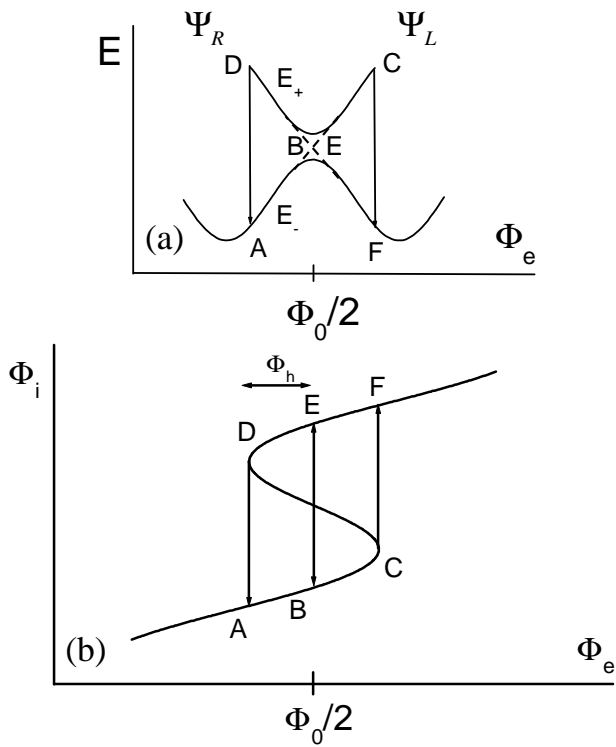}
\caption{(a)~Quantum energy levels of the 3JJ flux qubit vs external flux.
The dashed lines correspond to the classical potential minima. In
all graphs, the states A, B, C correspond to, say, $\Psi_L$ with
left-rotating supercurrent. As $\Phi_e$ is increased, these lose
classical stability in favour of the corresponding states
$\Psi_R$, denoted by D, E, and F. (b)~Internal vs external qubit
flux.} \label{fig1}
\end{figure}

In the basis $\{\psi_L,\psi_R\}$ and near
$\Phi_\mathrm{e}=\frac{1}{2}\Phi_0$, the flux qubit can be
described by the Hamiltonian
\begin{equation}\label{eq01}
  H = - {\Delta\over 2} \sigma_x - {\varepsilon \over 2} \sigma_z.
\end{equation}
At bias $\varepsilon = 0$, the two lowest levels of the qubit
anticross (Fig.~\ref{fig1}a) with energy gap $\Delta$. With
$\varepsilon$ changing sign, the qubit can either adiabatically transform
from $\psi_L$ to $\psi_R$ staying in the ground state $E_-$ or
switch to the excited state~$E_+$. The probability of the latter
(called Landau-Zener transition) for linear sweep
$\varepsilon(t) = \nu t$ and $\varepsilon$ changing from $-\infty$ to
$\infty$ was calculated~\cite{LZ} to be
$P_\mathrm{LZ}=\exp(-\pi\Delta^2\!/2\hbar\nu)$.

In order to demonstrate the principle of the IMT measurements of
this system, let us consider the internal flux representation
(Fig.~\ref{fig1}b) instead of the energy one (Fig.~\ref{fig1}a). A
similar picture is usually used for the explanation of the
operation of the conventional radio-frequency (rf) SQUID. The main
difference between an rf SQUID and a qubit behavior is the
existence of the adiabatic trajectory $BE$ for latter one (see
Fig.~\ref{fig1}a, b). Let us assume that $BE$ trajectory is
forbidden and ``qubit'' is inductively coupled to the high-quality
resonant circuit. Then the system exhibits hysteretic
behavior.~\cite{ili02} The tank circuit is simultaneously driven
by a dc bias current $I_{dc}$ and an ac current $I_{rf}$ of a
frequency $\omega$ close to the resonance frequency of the tank
circuit. Both currents produce the total  magnetic flux applied to
the qubit $\Phi_e = \Phi_{dc}+\Phi_{rf}\cos \omega t$. If the
amplitude $\Phi_{rf}>\Phi_{h}$, where $\Phi_{h}$ is the half-width
of the hysteresis loop $ACFD$ (Fig.~\ref{fig1}b), the tank circuit
will register the energy losses proportional to the loop area, as
long as $|\Phi_{dc}-\frac{1}{2}\Phi_0|<\Phi_{rf}-\Phi_{h}$. These
losses occur due to the jumps from $E_+$ to $E_-$ at the ends of
the loop. This idea was used by A. Silver and J. Zimmerman to
build the first rf SQUID magnetometers.~\cite{Zimm} If
$\Phi_{rf}>\Phi_{h}$ the rf voltage across the tank circuit is a
$\Phi_0$-periodical function of applied dc flux $V_T(\Phi_{dc})$
with local minima at $\Phi_{dc}=\frac{1}{2}\Phi_0 + n\Phi_0$,
where $n$ is integer.
\begin{figure}
\includegraphics[width=8.1cm]{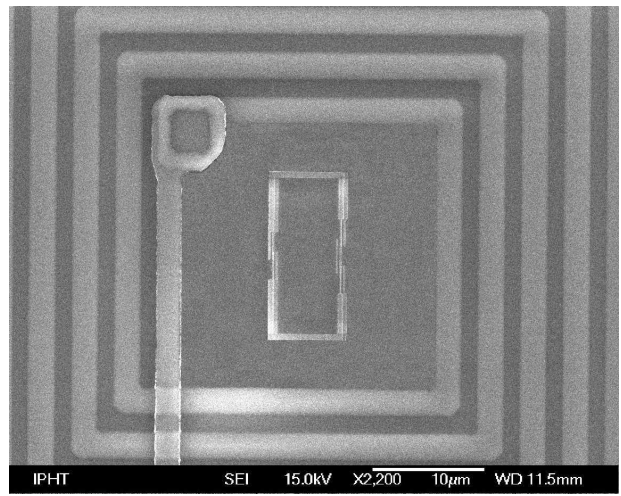}
\caption{Electron micrograph of the qubit at the center of the
tank coil.} \label{fig2}
\end{figure}

Now, let us take into account the additional ``quantum'' trajectory $BE$ (see
Fig.~\ref{fig1}b). If its probability $1-P_{LZ}$
is nonzero but less than $1$, two new closed
paths  $ABED$ and $BCFE$ are possible. There are two contributing
trajectories,   adiabatic and Landau-Zener transition.
Therefore the net dissipation is $P_{loss} = 2P_{LZ}(1-P_{LZ})$
and vanishes if $P_{LZ}$ is either too small or too large \cite{zen}.
Due to  the exponential dependence of $P_{LZ}$ on the sweep
rate, in practice this makes the $quantum$ losses observable only if the bias sweep narrowly
overshoots the anticrossing, i.e., if
\begin{equation}
|\Phi_{dc}-\frac{1}{2}\Phi_0|\lesssim \Phi_{rf},
\label{Eq:dipcond}
\end{equation}
when $\Phi_e$ changes slowly. Plotting $V_T(\Phi_{dc})$ for
$\Phi_{rf}>\Phi_{h}/2$, a plateau flanked by two peaks is expected.
The position of the dips depends on $\Phi_{rf}$ as follows from
Eq.~\ref{Eq:dipcond}.
Therefore in contrast to $V_T (\Phi_{dc})$ dependence of an rf SQUID
the qubit should exhibit two local minima (in one
period) which are symmetrical with respect to
$\Phi_{dc}=\frac{1}{2}\Phi_0$. For the amplitude
$\Phi_{rf}>\Phi_{h}$ the $ACFD$ hysteresis becomes closed as well.
Here, similar to the rf SQUID, on the $V_{rf}(\Phi_{dc})$ dependence
should appear the local minimum exactly at
$\Phi_{dc}=\frac{1}{2}\Phi_0$. Note, that $\Phi_e$ here plays a
role of bias $\varepsilon$ for the Hamiltonian (\ref{eq01}).

To test the ideas discussed above, we prepared lithographically
square-shaped Nb pancake coils with inductance $L_T$ on oxidized
Si substrates for the tank circuits. We used an external
capacitance $C_T$ to be able to change the resonant frequency
$\omega_T = 1/\sqrt{L_TC_T}$. The line width of the 30 coil
windings was 2~$\mu$m, with a 2~$\mu$m spacing. The quality factor
of the tank was $Q_T \approx 1500$ at $\omega_T \sim 20$~MHz.
\begin{figure}
\includegraphics[width=8.1cm]{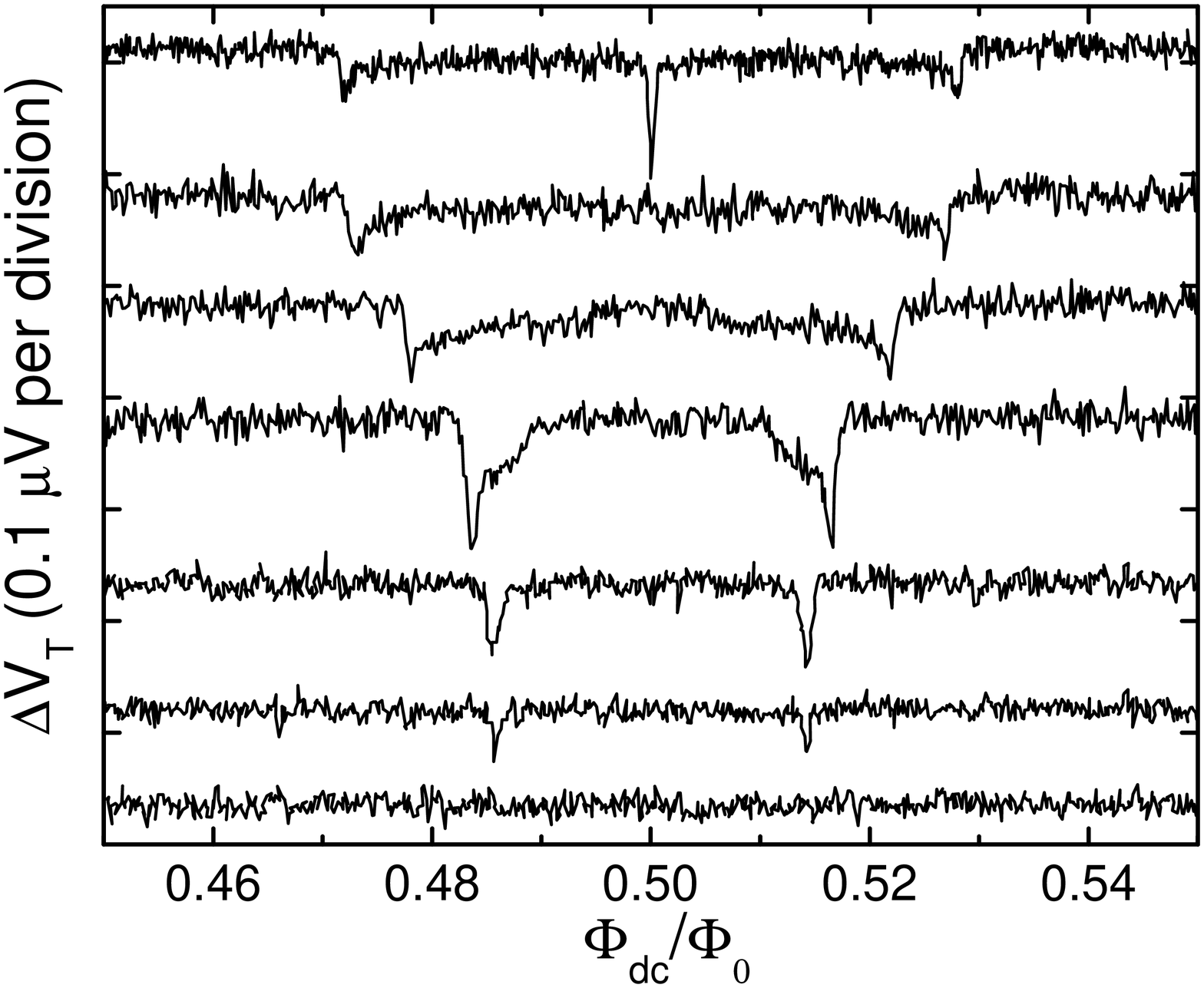}
\caption{Tank voltage vs. magnetic flux bias near
the degeneracy point of the qubit
$\frac{1}{2}\Phi_0$. From the lower to upper curve, the driving
voltage is 10.2, 10.7, 11.2, 13.1, 17.2, 21.3, 22.0 $\mu$V rms
(data vertically shifted for clarity).} \label{fig3}
\end{figure}
The 3JJ qubit structure was fabricated out of Al in the middle of
the coil by conventional shadow evaporation technique. The
Josephson junctions with critical current density $j_c\approx
300$A/cm$^2$ have areas $\approx130\times620$~nm$^2$,
$120\times600$~nm$^2$, and $110\times610$~nm$^2$ respectively.
The loop area was 90~$\mu\mathrm{m}^2$, with $L_q =
39$~pH. The fabricated structure is shown in Fig.~\ref{fig2}.

We measured $V_T (\Phi_{dc})$ by a three-stage cryogenic amplifier
placed at $\approx2$~K.~\cite{Oukhansky02} Results for small
driving voltage are shown in Fig.~\ref{fig3}. For the smallest
voltages no dissipative response is observed; the two ``quantum''
peaks appear around $10.7~\mu$V,~\cite{zen} and subsequently move
apart without significant broadening. The ``classical'' peak
appears in the center, and with an ac bias threshold \emph{double}
the one of the quantum peaks---both as predicted above.

Now assume that the probability of a Landau-Zener
transition is small and the qubit adiabatically changes from
$\psi_L$ to $\psi_R$, always staying in the ground state~$E_-$.
This means that the hysteresis $ACFD$ is ``shunted'' by the $BE$
trajectory. Therefore there are no losses caused by the flux jumps
in the qubit. However, in the vicinity of $B$ (see
Fig.~\ref{fig1}b) the small change of the external magnetic flux
causes a considerable change of the flux inside the qubit. Due to
coupling of the qubit to the tank, the effective inductance of the
tank-qubit system is changed, which leads to the change of the
resonant frequency. In this context the convenient measure of that
change is the imaginary part of the total
impedance~\cite{Ilichev01} expressed through the phase angle
$\chi$ between driving current $I_{bias}(t) = I_{ac}\cos \omega t
,$ and tank voltage $V_T(t) = V_T \cos
(\omega t + \chi)$. For small~$L_q$ and if the amplitude of
$I_{rf}$ is negligible, the results are summarized
by~\cite{Greenberg02b}
\begin{equation}\label{v}
  \tan\chi=k^2Q_{T}L_q \frac{d^2E_-(f_x)}{d\Phi_e^2}\;,\displaybreak[0]\\
\end{equation}
where $k=M/\sqrt{L_q L_\mathrm{T}}$ is the tank--qubit coupling
coefficient, with $M$ being the mutual inductance between them.
The ground-state curvature is~\cite{Greenberg02b}
\begin{equation}\label{Eq:dE}
  \frac{d^2 E_-}{d\Phi_e^2}=-\frac{E_\mathrm{J}^2\Delta^2\lambda^2}
    {\Phi_0^2(4E_\mathrm{J}^2\lambda^2f_x^2+\Delta^2)^{3/2}}\;,
\end{equation}
where
\begin{equation}\label{Eq:f}
 f_x=\frac{\Phi_e}{\Phi_0} - \frac{1}{2},
\end{equation}
\begin{figure}
\includegraphics[width=8.1cm]{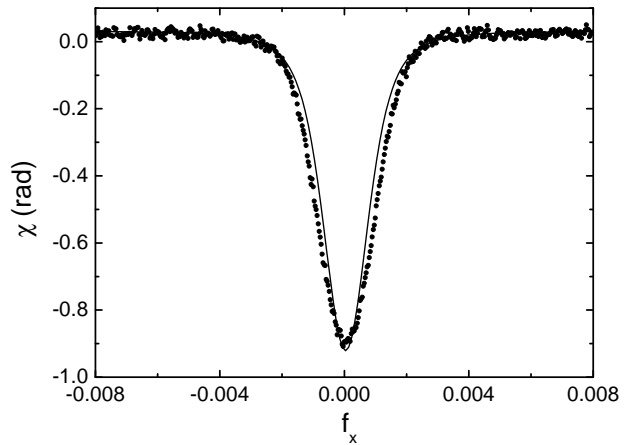}
\caption{Tank phase shift vs. flux bias near degeneracy $f_x~=~0$.
Dots correspond to experimental data, solid line is the
theoretical fit with $\Delta/h=650$~MHz.} \label{fig4}
\end{figure}
and $\lambda$ is the normalized flux-to-energy conversion factor.
Since all quantities in Eqs.~(\ref{v})--(\ref{Eq:f}) can be
measured independently, experimental results can be compared with
theoretical expectations.~\cite{miro}

Strictly speaking, Eq.~(\ref{v}) describes a measurement of the
quantum object with vanishing back-action. Therefore, its validity
should be proved.~\cite{SmirnovRS} Taking into account the
influence of the tank on the qubit, the Hamiltonian (\ref{eq01})
should be rewritten as:
\begin{equation}\label{H}
H = - {\Delta\over 2} \sigma_x - {\varepsilon \over 2} \sigma_z -
\sigma_z ( Q_0 + f + \gamma \hat{I}_T ) + H_T + H_{qB},
\end{equation}
where $\gamma = I_q M$ is the coupling coefficient between the
qubit's current, $\hat{I}_q = I_q \sigma_z,$  and the current in
the tank $\hat{I}_T$. An infinitesimally small auxiliary force
$f(t)$ is required for calculations of qubit's magnetic
susceptibility. A heat bath operator
$Q_0$ and a Hamiltonian $H_{qB}$ describe internal
mechanisms of dissipation and fluctuations in the superconducting
loop. The high-quality tank is treated here as a quantum cavity is characterized
by creation/annihilation photon operators $a^+,a$ which obey the
Bose commutation rules $[a,a^+]_- = 1$ etc. Quantum-mechanical
operators of a current and a voltage in the tank are defined as:
$\hat{I}_T = \sqrt{\hbar \omega_T/2 L_T}(a + a^+), \hat{V}_T =
i\sqrt{\hbar \omega_T/2C_T} (a^{+} - a)$. For the Hamiltonian of
the tank driven by a bias current $I_{bias}$ and coupled to its
own heat bath $Q_b$  we get the expression
\begin{eqnarray}
H_T &=& \hbar \omega_T (a^+a + 1/2) - (a + a^+)Q_b \notag\\
& &- L_T \hat{I}_T I_{bias} + H_{TB}.
\end{eqnarray}
The internal heat bath of the tank $Q_b$, characterized by a free
Hamiltonian $H_{TB}$, results in a finite life time of the
photons, $\gamma_T^{-1}$, and, because of this, in a finite
quality factor, $Q_T = \omega_T/\gamma_T.$ Assuming that $\hbar=1,
k_B=1, $ we derive the Heisenberg equations for the tank
operators: $\dot{\hat{I}}_T = \hat{V}_T/L_T, $ and
\begin{equation}
\left( {d^2\over dt^2} + \gamma_T {d \over dt} + \omega_T^2\right)
\hat{V}_T = \xi_b + \lambda \omega_T^2 \dot {\sigma}_z + {1\over
C_T} \dot{I}_{bias},
\end{equation}
where $\xi_T(t)$ is a fluctuation source with zero average value,
$\langle \xi\rangle = 0$, and a correlator, $\langle
\xi_b(\omega)\xi_b\rangle,$  which is proportional to the
linewidth of the tank $\gamma_T$ and the tank temperature T:
$\langle \xi_b(\omega)\xi_b\rangle =  (2\gamma_T T/C_T) \omega^2.$
Because of inductive coupling the current and the voltage in the
tank, $\hat{I}_T$ and $\hat{V}_T$, affect the qubit current,
$\hat{I}_q = I_q \sigma_z.$ Using the linear response
theory this influence can be presented as follows
\begin{equation}\label{sig}
 \dot{\sigma}_z  =  \dot{\sigma}_{z,0} +
{\lambda \over L_T} \int dt_1 \langle{\delta \sigma_z(t) \over
\delta f(t_1)}\rangle \hat{V}_T(t_1)
\end{equation}
where an operator $\sigma_{z,0}$ describes fluctuations of the
qubit current caused by its internal heat bath, $Q_0,$ which is
not correlated with the heat bath of the tank, $Q_b$. We also take
into account relations $(\delta /\delta I_T) = \lambda
(\delta/\delta f)$ and $\dot{\hat{I}}_T = \hat{V}_T/L_T. $

The function $\langle \delta \sigma_z(t) / \delta f(t_1)\rangle $
involved in Eq.~(\ref{sig}) is proportional to the derivative of
the qubit current $I_q(t) = \langle \hat{I}_q \rangle $ over the
flux $\Phi_T = L_T I_T$ created by the tank, $ \delta
I_q(t)/\delta \Phi(t_1),$ or to the second derivative of the qubit
energy profile, $E(\Phi),$ over the flux, $\partial^2
E(\Phi)/\partial \Phi^2$ (compare with Eq.~\ref{v}). It is
convenient to characterize the qubit response on the action of the
tank by means of the magnetic susceptibility~\cite{ili03}
$\chi_{zz}(\omega)$ defined as
\begin{equation}\label{mag}
\langle {\delta \sigma_z(t) \over \delta f(t')}\rangle= \int
{d\omega \over 2 \pi} e^{-i\omega(t-t')} \chi_{zz}(\omega).
\end{equation}
Then, the voltage in the tank obeys the equation
\begin{eqnarray}\label{vol}
\int dt_1 \left[ \left( {d^2\over dt^2} + \gamma_T {d \over dt} +
\omega_T^2\right)\delta(t - t_1) \right.\notag\\
\left. - {\lambda^2 \over L_T}\omega_T^2 \langle {\delta
\sigma_z(t) \over \delta f(t_1)}\rangle \right]
 \hat{V}_T(t_1)  = \nonumber\\
 \xi_b + \lambda \omega_T^2 \dot{\sigma}_{z,0} +
 \lambda \omega_T^2
{1\over C_T} \dot{I}_{bias}.
\end{eqnarray}
It is evident from this equation that the tank voltage contains
information about the magnetic susceptibility $\chi_{zz}(\omega)$
of the qubit. Similar to classical case this information can be
extracted from measurements of the phase angle $\chi$. It follows
from the averaged Eq.~(\ref{vol}) that amplitudes of harmonic
oscillations of the tank voltage and the bias current are related
through
\begin{eqnarray}
V_T e^{-i\chi} &=& - i \omega \left\{   \omega_T^2\left[ 1 -
{\lambda^2 \over L_T} \chi_{zz}^{\prime }(\omega)\right] -
\omega^2 \right. \notag\\
 & &\left. - i \omega \left[ \gamma_T + {\lambda^2 \omega_T^2\over
\omega L_T} \chi_{zz}^{\prime\prime}(\omega )\right] \right\}^{-1}
{I_{ac} \over C_T}
\end{eqnarray}
with  $\chi_{zz}^{\prime }(\omega)$ and $\chi_{zz}^{\prime\prime
}(\omega)$ being the real and imaginary parts of the qubit
magnetic susceptibility (\ref{mag}). In the case when the tank is
driven exactly with the resonant frequency, $\omega = \omega_T,$
the voltage amplitude, $V_T,$ can be found from the equation:
\begin{eqnarray}
V_T &=& {I_{ac} \over C_T} \left\{ [  k^2 L_q I_q^2 \omega_T
\chi_{zz}^{\prime}(\omega_T)]^2 \right. \notag\\
 & &\left. + [ \gamma_T + k^2 L_q I_q^2
\omega_T \chi_{zz}^{\prime\prime}(\omega_T)]^2  \right\}^{-1/2},
\end{eqnarray}
whereas for the voltage-current phase shift we obtain the
expression
\begin{equation}
\tan \chi = - 2 k^2 L_qI_q^2  \bar{Q}_T
\chi_{zz}^{\prime}(\omega_T) .
\end{equation}
Here $\bar{Q}_T = \omega_T/(\bar{\gamma}_T)$ is an effective
quality factor of the tank wherein a broadening of tank's line
width,
\begin{equation}\label{gam}
 \bar{\gamma}_T = \gamma_T + k^2 L_qI_q^2 \omega_T \chi_{zz}^{\prime\prime}(\omega_T ),
\end{equation}
 due to the qubit is taken into account.
The magnetic susceptibility of the qubit (Eq.~\ref{mag}) is
calculated from the Bloch equations written in the form:
\begin{eqnarray}\label{eq_Bloch}
\langle\dot{\sigma}_x\rangle + \Gamma_x ( \langle \sigma_x\rangle
- \sigma_x^0)
= - \varepsilon \langle \sigma_y \rangle,  \nonumber\\
\langle \dot{\sigma}_y \rangle + \Gamma_y \langle \sigma_y \rangle
= - \Delta \langle \sigma_z\rangle  +  \varepsilon \langle
\sigma_x \rangle  - 2 f \sigma_x^0, \nonumber\\
\langle \dot{\sigma}_z\rangle  = \Delta \langle \sigma_y\rangle ,
\end{eqnarray}
where $\Gamma_x$ and $\Gamma_y$ are qubit's damping rates,
$\sigma_x^0 = - (\Delta/\omega_c  ) \tanh(\omega_c /2T)$ is a
steady-state polarization of the qubit with energy splitting
$\omega_c =   \sqrt{\Delta^2 + \varepsilon^2} $, which is much
higher than the resonant frequency of the tank, $\omega_c \gg
\omega_T.$ Because of this the decoherence and relaxation rates
drop out of the expression for the magnetic susceptibility:
\begin{eqnarray}
&\chi_{zz}&(\omega_T) = \chi_{zz}'(\omega_T) = \notag\\
& &= 2\frac{\Delta^2}{(\Delta^2 + \varepsilon^2)^{3/2}}
\tanh\left( \frac{\sqrt{\Delta^2 + \varepsilon^2}}{2T}\right).
\end{eqnarray}
As a result, the phase angle between the voltage in the tank and
the bias current is given by the formula
\begin{eqnarray}\label{chi}
\tan \chi = - 2 k^2 \frac{L_q I_q^2}{\Delta} \bar{Q}_T
\left(\frac{\Delta^2}{\Delta^2 + \varepsilon^2} \right)^{3/2}\notag\\
\times\tanh\left( \frac{\sqrt{\Delta^2 +
\varepsilon^2}}{2T}\right).
\end{eqnarray}
By making use of simple algebra it can be shown that at $T=0$  Eqs.~(\ref{v})
and (\ref{chi}) are equivalent. Therefore, indeed measuring $\tan
\chi$ as a function of the bias applied to the qubit, let us to
determine the qubit's tunneling rate $\Delta.$

In order to realize the adiabatic response  of the qubit
experimentally, we fabricated 3JJ Al qubit with the following
parameters. The area of two, nominally equivalent junctions was
about 190x650~nm$^2$ while one is smaller, so that $\alpha\equiv
E_{\mathrm{J}3}/E_{\mathrm{J}1,2}\approx0.8$. The value of the
critical current for larger junction was determined to be
$I_\mathrm{c}\approx380$~nA. Qubit inductance, tank
parameters, and measurement setup were the same as in the case of
Landau-Zener transitions described above.

The measured $\chi(f_\mathrm{x})$ curve  at nominal mixing-chamber
temperature $T=10$~mK is shown in Fig.~\ref{fig4}. The curve was
fitted by Eq.~(\ref{v})) with $\Delta$ as a free parameter.
Calculated curve for the best fit parameter $\Delta/h=650$~MHz is
also shown Fig.~\ref{fig4}. This value of energy gap was in good
agreement with the gap determined independently from temperature
measurements.~\cite{miro}

\section{Rabi spectroscopy}
Quite generally, a two-level quantum system (including qubits),
should exhibit coherent (Rabi) oscillations in time in the
presence of resonant irradiation. Microwaves in resonance with the
spacing between qubit's energy levels will cause their occupation
probabilities to oscillate, with a frequency proportional to the
microwave amplitude. Indeed, coherent oscillations between energy
levels of the superconducting qubit were
detected.~\cite{NakamuraPashkinTsai,Esteve,Martinis,NakamuraDelft}

In this section we show that the tank can be used for detection of
Rabi oscillations as well. If the resonant microwave signal is
applied, the phase-coherent oscillations of the level occupation
will only last for a finite time, which is usually called the
coherence time. The correlation between the occupations can be
expressed by an autocorrelation function or its Fourier transform,
the spectral density. For example for the IMT, when the flux qubit
is coupled inductively to a tank circuit, the spectral density of
the tank-voltage fluctuations rises above the background noise
when the qubit's Rabi frequency~$\Omega_\mathrm{R}$ coincides with
the tank's resonant frequency~$\omega_\mathrm{T}$. This forms the
basis for our measurement technique of \emph{Rabi spectroscopy}.
Rabi oscillations cause changes of the qubit's magnetic moment
and, therefore, excite the tank. The tank circuit accumulates
photons which were emitted by the qubit.  This approach is similar
to the one in entanglement experiments with Rydberg atoms and
microwave photons in a cavity.~\cite{Raimond01}

Indeed, quantitative information can be extracted from the noise
spectrum $S_V(\omega)$ of the voltage fluctuations (the Fourier
transform of the correlator $ M_V(t,t') = \langle
(1/2)[\hat{V}_T(t),\hat{V}_T(t')]_+\rangle  $) in the
tank,~\cite{SmirnovRS} which incorporates the noise spectrum of
the tank, $S_{VT},$ supplemented by the qubit's contribution
$S_{VQ}, S_V = S_{VT}(\omega) + S_{VQ}(\omega), $ where
\begin{equation}
S_{VT}(\omega ) =  2 {\omega^2 \over C_T}  { T \gamma_T  \over (
\bar{\omega}_T^2  -  \omega^2 )^2 + \omega^2  \bar{\gamma}_T^2 }.
\end{equation}
The qubit's part of voltage noise can be found from the stochastic
equation (\ref{vol}) for the tank voltage:
 \begin{equation}
S_{VQ}(\omega ) =  \omega^2 {\omega_T\over C_T} \times  { k^2 L_q
I_q^2  \omega_T   S_{zz}(\omega )   \over (  \bar{\omega}_T^2  -
\omega^2 )^2 + \omega^2  \bar{\gamma}_T^2 }.
\end{equation}
Here $ S_{zz}(\omega )$ is a Fourier transform of the correlator
$\langle {1/2} \left[\sigma_{z,0}(t),
\sigma_{z,0}(t')\right]_+\rangle,$  which describes internal
fluctuations in the qubit (not related to the tank). Hand in hand
with the tank's damping rate, $\bar{\gamma}_T$ (\ref{gam}), the
resonance frequency of the tank, $ \bar{\omega}_T,$ is also
shifted because of the qubit-tank interaction,
\begin{equation}
\bar{\omega}_T  = \omega_T\sqrt{ 1 - k^2 L_q I_q^2
\chi_{zz}^{\prime }(\omega_T)}.
\end{equation}
The spectrum of voltage fluctuations has a peak near the resonant
frequency of the tank $\omega_T,$ and, therefore, it contains
information about a low-frequency component  $S_{zz}(\omega_T)$ of
the qubit spectrum. The equilibrium part of this spectrum peaks at
the energy splitting $\omega_c = \sqrt{\Delta^2 + \varepsilon^2}$
of the tunneling doublet which differs significantly from the
frequency of the tank, $\omega_c \gg \omega_T.$ Because of this, a
contribution of equilibrium qubit noise to the voltage spectrum of
the tank is expected to be negligibly small. An external
microwave source with a frequency $\omega_0$ tuned in resonance
with the energy splitting of the qubit $\omega_c$, induces
periodical variations of a population difference between the
excited and ground state of the qubit, which are characterized by
a frequency $\Omega_R = \sqrt{(\Delta /\omega_c)^2 F^2 +
\delta_0^2} $ depending on the amplitude $F$ of the microwave
source as well as on the detuning $\delta_0 = \omega_0 - \omega_c.$
With non-zero bias, $\varepsilon \neq 0,$ left and right wells of
the qubit potential have different energies. As a consequence, Rabi
oscillations between the energy eigenstates will be accompanied by
low-frequency transitions of the qubit from the left  to the right
well and back. The tank detects this kind of low-frequency noise
which is described by the Lorentzian spectrum centered at the Rabi
frequency $\Omega_R$ with a linewidth dependent on the qubit
decoherence rate $\Gamma$. Both the tank  ($\Gamma_T$) and the
internal heat bath ($\Gamma_0$) contribute to the decoherence rate
$\Gamma, \Gamma= \Gamma_0 + \Gamma_T.$ It should be emphasized
that the external microwave field affects the qubit-bath coupling \cite{SmirnovDeco}
that results in the distinction of the nonequlibrium decay rate
$\Gamma$ from its equilibrium counterparts $\Gamma_x, \Gamma_y$
entering the Bloch equations (\ref{eq_Bloch}).

An informative part of the spectrum of voltage fluctuations,
$S_{VQ}(\omega),$ incorporates the qubit Lorentzian  multiplied by
the transmission function of the tank having a sharp peak at the
frequency $\omega_T$ :
\begin{eqnarray}\label{Gtot}
S_{VQ}(\omega ) = 2 \frac{\varepsilon^2}{\omega_c^2} k^2 \frac{L_q
I_q^2}{C_T} \omega^2 \Gamma_0  \frac{\omega_T^2}
{(\bar{\omega}_T^2  -  \omega^2)^2 + \omega^2  \bar{\gamma}^2_T }\notag\\
\times  \frac{ \Omega_R^2} {(\omega^2 - \Omega_R^2)^2 + \omega^2
\Gamma^2 }.
\end{eqnarray}
The linewidth of the tank is assumed to be much less than the qubit's
damping rate, $\gamma_T \ll \Gamma.$  Because of this, as a
function of frequency $\omega$, the spectrum of voltage noise (\ref{Gtot})
represents a Lorentzian with a width $\gamma_T$ and an amplitude
which is given by a Lorentzian function of the Rabi frequency having
the maximum near $\omega_T$ and the width $\Gamma.$
Measurements of the noise spectrum amplitude at different values
of the microwave power $P$ allow to extract information not only
about the existence of Rabi oscillations, but also about the
nonequilibrium decoherence rate $\Gamma$ of the qubit.  We note
that due to strong nonequilibrium conditions the populations of
the qubit's levels are practically equal, and the noise
spectrum amplitude does not depend on the temperature. The
signal-to-noise ratio,
\begin{eqnarray}
{ S_{VQ}(\omega )\over S_{VT}(\omega )}_{|\omega = \omega_T} =
{\varepsilon^2 \over \omega_c^2} k^2 {L_q I_q^2 \over T} {\Gamma_0
\over \gamma_T} \notag\\
\times{ \omega_T^2 \Omega_R^2  \over (\omega_T^2 - \Omega_R^2)^2 +
\omega_T^2 \Gamma^2 },
\end{eqnarray}
peaks when $\Omega_R = \omega_T.$ At the same point, the backaction
of the measuring device (tank) on the quantum bit which is
described by the damping rate $\Gamma_T,$
\begin{equation}\label{S}
\Gamma_T = 4 k^2 L_q I_q^2 {\varepsilon^2 \over \omega_c^2}
\omega_T^2
 { T \gamma_T \over
(\omega_T^2 - \Omega_R^2)^2 + \Omega_R^2 \gamma_T^2 }.
\end{equation}
reaches its maximum as well. However, the tank contribution to the
qubit decoherence drastically decreases with small detuning of the
Rabi frequency $\Omega_R$ from $\omega_T \gamma_T \ll |\Omega_R -
\omega_T| < \Gamma.$ At the same time, the efficiency of
measurements, $( S_{VQ}(\omega )/S_{VT}(\omega ))_{|\omega =
\omega_T},$ remains practically unchanged. Since $\Gamma \ll
\Omega_R, $, the spectroscopic monitoring of Rabi oscillations
with the low-frequency tank circuit falls into the category of
weak continuous quantum measurements.

\begin{figure}
\includegraphics[width=8.1cm]{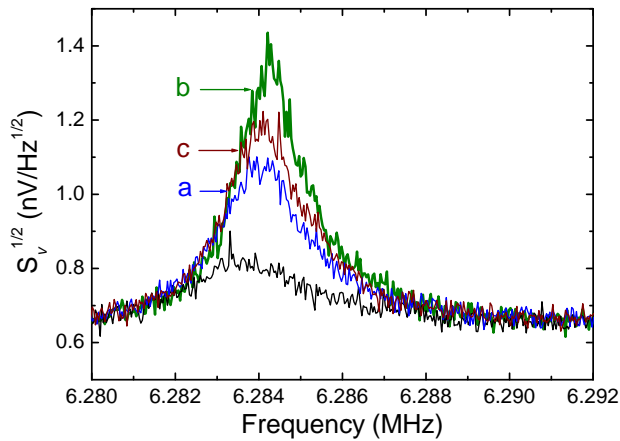}
\caption{The spectral noise amplitude of the tank voltage for UHF powers
$P_a <P_b <P_c$ at 868 MHz. The bottom curve corresponds to the
background noise without an HF signal.} \label{fig5}
\end{figure}

The measurement setup as well as sample fabrication were similar
to the ones described in the previous section. Microwave irradiation
(UHF signal) was
introduced to the sample by a commercial coaxial cable (between
room temperature and $\sim$~2~K) and the resistive coaxial cable,
known as ThermoCoax (between $\sim$~2~K and 10~mK). In order to
reduce external disturbances, a 20~dB commercial attenuator
was installed at 2~K. To measure
$S_V$, we tuned the UHF signal in resonance with the qubit level
separation. We found noticeable output signal only when
$\omega_{_{\rm HF}}/2\pi=868\pm2$MHz, in agreement with the
estimated splitting $\Delta/h\sim1$GHz. Note that there is a
difference of two orders of magnitude between $\omega_{_{\rm HF}}$
and the readout frequency~$\omega_{T}$.
Together with the high $Q_{T}$, this ensures that the signal can
only be due to resonant transitions in the qubit itself. This was
verified by measuring $S_V$  when biasing the qubit away from
degeneracy. A signal exceeding the background, that is, emission
of $\sim$~6~MHz photons by the qubit in response to a resonant UHF
field in agreement with Eq.~(\ref{S}), was only detected when the
qubit states were almost  degenerate (cf.\ below Eq.~(\ref{Gtot})).
The measurements were carried out at nominal temperature
$T=10$~mK. No effect of radiation was observed above 40~mK (with
40~mK$/hk_\mathrm{B}\approx830$~MHz, i.e.\ close to~$\Delta/h$)
\begin{figure}
\includegraphics[width=3.25in]{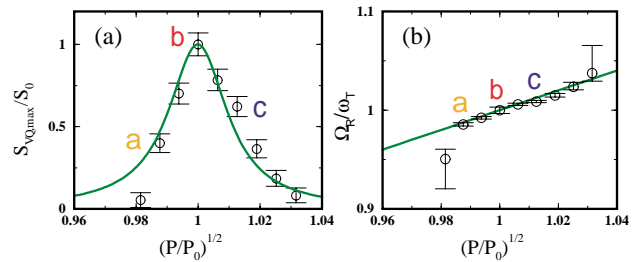}
\caption{(a)~Comparing the data to the theoretical Lorentzian. The
fitting parameter is $g\approx0.02$. Letters in the picture
correspond to those in Fig.~\ref{fig5}. (b)~The Rabi frequency
extracted from~(a) vs the applied UHF amplitude. The straight line
is the predicted dependence
$\omega_\mathrm{R}/\omega_\mathrm{T}=\sqrt{P/P_0}$. The good
agreement provides strong evidence for Rabi
oscillations.}\label{fig6}
\end{figure}
We plotted $S_{V}(\omega)$ for different HF powers $P$ in
Fig.~\ref{fig5}. As $P$ is increased, $\omega_\mathrm{R}$ grows
and passes~$\omega_\mathrm{T}$, leading to a non-monotonic
dependence of the maximum signal on~$P$ in agreement with the
above picture. This and the sharp dependence on the tuning of
$\omega_{_{\rm HF}}$ to the qubit frequency confirm that the
effect is due to Rabi oscillations.

For a quantitative comparison between theory and experiment, we
subtracted the measured signal without an HF power from the
observed $S_{V}$, yielding the qubit's contribution $S_{VQ} = S_V
- S_{VT}(\omega). $ Subsequently, we extracted the peak values vs.
UHF amplitude,
$S_{VQ,\mathrm{max}}(\sqrt{P/P_0})=\max_{\omega}S_{VQ}(\omega)\approx
S_{VQ}(\omega_\mathrm{T})$, where $P_0$ is the power causing the
maximum response; see Fig.~\ref{fig6}a. In the same figure, we
plot the theoretical curve for $S_{VQ,\mathrm{max}}$ normalized to
its maximum~$S_0$,
\begin{equation}\label{S_norm}
 \frac{S_{VQ,\mathrm{max}}(w)}{S_{0}}  = \frac{w^2 g^2}{(w^2{-}1)^2 + g^2}
  \approx \frac{(g/2)^2}{(w{-}1)^2 + (g/2)^2}\;;
\end{equation}
$w\equiv\Omega_\mathrm{R}/\omega_\mathrm{T}$ ($=\sqrt{P/P_0}$
theoretically) and $g=\Gamma/\omega_\mathrm{T}$. The best fit is
found for
$\Gamma\approx0.02~\omega_\mathrm{T}\sim8\cdot\nobreak10^5~\mathrm{s}^{-1}$.
Thus, the life-time of the Rabi oscillations is at least
$\tau_\mathrm{Rabi}=2/\Gamma\approx2.5~\mu$s, leading to an
effective quality factor
$Q_\mathrm{Rabi}=\Delta/(\hbar\Gamma)\sim7000$. These values
substantially exceed those obtained recently for a modified 3JJ
qubit ($\tau_\mathrm{Rabi}\sim150$ns),~\cite{NakamuraDelft} which
is not surprising. In our setup the qubit is read out not with a
dissipative DC-SQUID, but with a high-quality resonant tank. The
latter is weakly coupled to the qubit ($k^2\sim10^{-3}$),
suppressing the noise leakage to it.~\cite{Rabi}

\section{Resonant spectroscopy} In this section we show that the IMT
can be also used for resonant spectroscopy, which is a well-known
experimental method for investigation of quantum systems. As an
example of such IMT application let us consider an
interferometer-type charge qubit.~\cite{Free,Zorin,Krech} The
device's core element is a single-Cooper-pair transistor - a small
island, separated by two mesoscopic Josephson junctions, which is
capacitively coupled to the gate. The transistor can be described
by the Hamiltonian matrix \cite{av,Tinkham}
\begin{equation}\label{Eq:HamMat}
  H_{nm}=4E_C(N-n_g)^2\delta_{nm}-\frac{\varepsilon_J(\varphi)}{2}
  (\delta_{n,m+1}+\delta_{n,m-1})
\end{equation} where $N$ is number of Cooper pairs on the island,
$\delta_{n,m}$ is Kronecker symbol, $E_C=e^2/2C_\Sigma$ is
the single-electron charging energy expressed through the total island
capacitance $C_\Sigma$. The dimensionless parameter $n_g=C_gV_g/2e$ is
continuously controllable by the gate voltage $V_g$ via the
capacitance $C_g$.  The effective Josephson energy
\begin{equation}\label{jos}
  \varepsilon_{J}(\varphi)=[E_{J1}^{2}+E_{J2}^{2}+2E_{J1}E_{J2}\cos\varphi]^{1/2},
\end{equation} is a function of the total phase difference across both
junctions $\varphi=\phi_1+\phi_2$, where $E_{J1,J2}$ and $\phi_{1,2}$
are Josephson coupling energies and phase differences of the first and
second junction, respectively.
\begin{figure}
\includegraphics[width=3in]{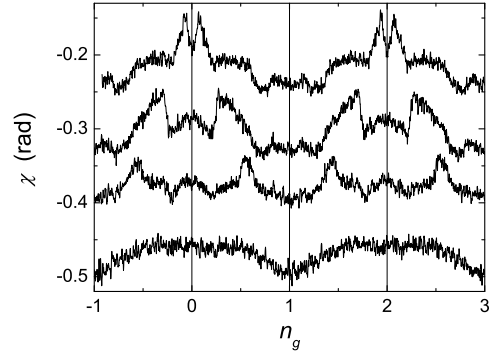} \caption{Tank phase shift $\chi$
vs. gate parameter $n_g$ without microwave power (lowest curve) and
with microwave power at different excitation frequencies. The data
correspond to $\omega_{UHF}/2\pi$ = 8.9, 7.5, 6.0 GHz (from top to
bottom). The magnetic flux $\Phi_e=\Phi_0/2$ threading the
interferometer loop provides a total phase difference $\delta=\pi$
across the single-Cooperpair transistor. (For clarity, the upper
curves are shifted.)} \label{fig7}
\end{figure}

\begin{figure}
\includegraphics[width=3in]{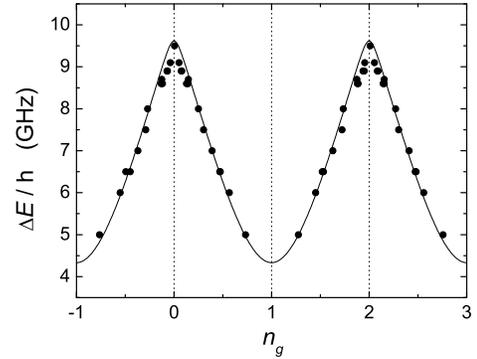}
\caption{Energy gap $\Delta$ between the ground and upper bands of
the transistor determined from the experimental data for the case
$\delta=\pi$. Some examples of these data are shown in
Fig.~\ref{fig7}. Dots represent the experimental data, the solid
line corresponds to the fit (cf. text).} \label{fig8}
\end{figure}

If the transistor is closed by a superconducting loop with low
inductance $L_q$, the total phase difference is $\phi\approx
2\pi\Phi_e/\Phi_0$ and the ground-state curvature
$d^2E_-/d\Phi_e^2$ can be obtained finding smallest eigenvalue of
the Hamiltonian matrix (\ref{Eq:HamMat}) as a function of
$\Phi_e$. Using (\ref{v}), we can calculte the phase shift of the
tank  inductively coupled to the charge qubit  and compare it with
experimental results obtained by IMT.~\cite{born} The principle of
resonant spectroscopy is very simple. If the qubit is irradiated
by microwaves with frequency coresponding to the energy gap
between ground ($n$=0) and upper level ($n$=1) the latter one
becomes also populated. In this sense the microwave irradiation
acts like temperature i.e. suppresses the tank phase shift (see
Eq.\ref{chi}).

Similar to the phase qubits, the interferometer-type charge qubit
was fabricated out of Al by the conventional shadow evaporation
technique, and was placed in the middle of the Nb coil by making
use of a flip-chip configuration. The geometric loop inductance of
the interferometer was calculated to be $L_q=0.8$~nH. The layout
size of the junctions was 140~nm x 180~nm. Deviations from the
nominal dimensions caused by the fabrication process were
estimated from the micrograph of the real structure and found to
be less than 15~$\%$. The charging energy was overestimated within
the framework of the plate capacitor model from the junctions
delivering $E_C \simeq 7$~GHz. In fact and also in accordance with
the experimental results below, this value is reduced due to the
strong tunneling regime.~\cite{zaik} The measurements were
performed at mixing chamber temperature of 10~mK.

The presence of the microwave power significantly changes the
obtained dependence, namely peaks appear in the $\chi(n_g)$ curve
(see upper curves in Fig.~\ref{fig7}). The peak position depends on
the microwave frequency and does not depend on the
amplitude(shape slightly depends). These peaks
disappear when the phase bias is far from $\pi$ as well as at
higher temperatures. Therefore, we believe, that they correspond
to the excitation of the system from the ground to the upper
state.

The microwave induced transition (both the
frequency of the microwave and the phase difference across the
transistor  $\varphi=\pi$ are fixed) from the ground to the upper
state occurs only at certain value of the gate charge. From the
position of the peaks on the $\chi(n_g)$ curves at different
frequencies of the microwave, we have reconstructed the energy
difference between ground and upper states as a function of the
quasicharge on the island. The obtained dependence is shown in
Fig.~\ref{fig8}. We fitted the experimental data by using the
numerical solution of the energy spectrum of the Hamiltoniani matrix
(\ref{Eq:HamMat}). The best fitting parameters were found to be
$\varepsilon_J(\pi)
= 4.4$~GHz and $E_C = 2.2$~GHz. This value of the Josephson
coupling energy is in very good agreement with the estimated
value, and, as expected, the charging energy is smaller than
estimated.

\section{Nonresonant spectroscopy of two coupled qubits}
After the successful demonstration of quantum coherence in many
types of superconducting qubits an observation of entangled states
in two coupled qubits presents the next step on the road to the
quantum processor. The entangled states were recently observed in
both the charge~\cite{Pashkin} and the current-biased Josephson
junction~\cite{Berkley} qubits. In this section we demonstrate
that entangled states in a system of two inductively coupled flux
qubits~\cite{Majer} can be detected by  the IMT.~\cite{IzmalkovDQ}

\begin{figure}
\includegraphics[width=3in]{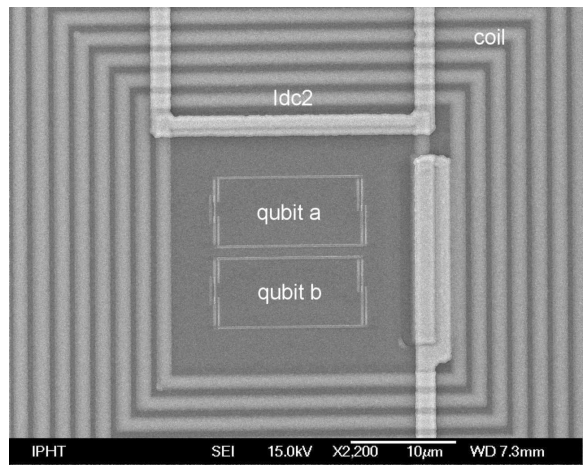}
\caption{Micrograph of the two-qubit system coupled to a resonant
tank circuit.} \label{fig9}
\end{figure}

The system of two flux Al qubits inductively coupled to each other
and to the Nb tank is shown in Fig.~\ref{fig9}. The area of each
qubit and self-inductance, and critical current were $S_q =
80$~$\mu$m$^2$, $L_q = 39$~pH, $I_c \approx 400$~nA, and $E_C
\approx 3.2$~GHz, respectively. The mutual inductance between the
qubits $M_{ab}=2.7$~pH was estimated numerically from the electron
micrograph. The magnetic flux through the qubits was created by
the dc component of the current in the coil $I_{dc1}$ and by the
bias current $I_{dc2}$ through a wire close to one of the qubits.
This allowed independent control of the bias in each qubit.
The system  of Fig.~\ref{fig9} is described by the Hamiltonian $H
= H_0 + H_T + H_{int} + H_{diss}$, where the two-qubit Hamiltonian
in the two-state approximation is expressed as~\cite{Alec03}
\begin{equation}\label{eq_Hamiltonian_0} H_0 =-\Delta_a
\sigma_{x}^{(a)} - \Delta_b \sigma_{x}^{(b)} + \epsilon_a
\sigma_{z}^{(a)} + \epsilon_b \sigma_{z}^{(b)} + J
\sigma_{z}^{(a)} \sigma_{z}^{(b)},
\end{equation}
$H_T$ is the tank Hamiltonian (a harmonic oscillator), the
qubit--tank interaction is
\begin{equation}\label{eq_Hamiltonian_int}
H_{int}=-(\lambda_a \sigma_{z}^{(a)} + \lambda_b \sigma_{z}^{(b)}
)I_T,
\end{equation}
and $H_{diss}$ describes the standard weak coupling of the qubits
to a dissipative bath~\cite{Weiss}.
\begin{figure}
\includegraphics[width=3in]{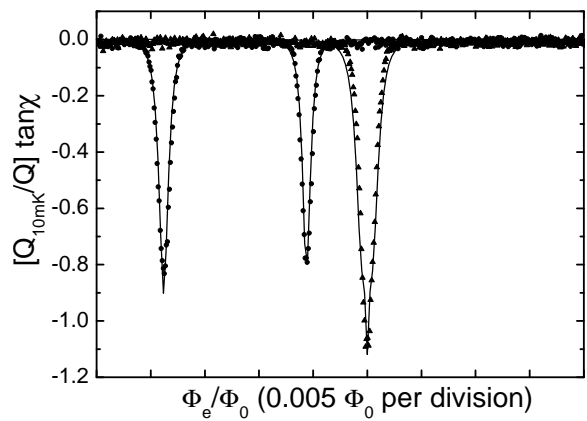}
\caption{Normalized tangent of the current-voltage angle $\chi$ in
the tank vs. external flux bias $\Phi_e$ at $50$~mK. A relative flux
bias $f_{shift}$ between the qubits is created by changing the
current $I_{dc2}$ in the additional wire. The shifted curves
correspond to $I_{dc2}$ = 27.3~$\mu$A, while the central curve is
for $I_{dc2}$ = -2.7~$\mu$A. The experimental data are denoted by
the dots ($I_{dc2}$ = 27.3~$\mu$A) and triangles ($I_{dc2}$ =
-2.7~$\mu$A). Solid curves correspond to theoretical fitting. }
\label{fig10}
\end{figure}

Here the coefficients are $\lambda_{a/b} =  M_{a/b, T} I_{a/b}$,
where $M_{a/b, T}$ is the qubit--tank mutual inductance, $L_{a/b}$
is the self-inductance and $I_{a/b}$ is the amplitude of the
persistent current in the corresponding qubit. In the standard
two-state approximation, the qubit current operators are
$\hat{I}_{a/b} = I_{a/b} \sigma_{z}^{(a/b)}$. The qubit biases are
given by $\epsilon_a = I_{a}\Phi_0 (f_x - 0.5 + f_{\rm shift})$,
$\epsilon_b = I_{b}\Phi_0 (f_x - 0.5 + \eta f_{\rm shift} )$,
where the dimensionless flux $f_x \sim I_{dc1}$ describes the
field created by the niobium coil in both qubits, while the
parameters $f_{\rm shift}\sim I_{dc2}$ and $\eta = M_{bw}/M_{aw} <
1$ give the bias difference between the qubits created by the
additional wire. Here $M_{aw}$ ($M_{bw}$) are the mutual
inductances between the $a$ ($b$) qubit and the additional dc wire
(for our sample, $M_{aw}$ and $M_{bw}$ were calculated
numerically, yielding $\eta=0.32$). The qubit--qubit coupling
constant $J = M_{ab} I_{a}I_{b}$ is positive because the two
qubits are in the same plane side to side, leading to
antiferromagnetic coupling (according to the north-to-south
attraction law).

The application of the IMT for spectroscopy of two coupled qubits,
similar to single-qubit problem (see section II), can be
conveniently discussed in terms of their magnetic susceptibility
$\chi_{zz}$. In the linear-response approximation the magnetic
susceptibility of the two-qubit system $\chi_{zz}(\omega )$ is
expressed through retarded Green's functions of the qubit
operators $\sigma_z^{(a/b)}$, calculated with the equilibrium
density matrix $\rho = e^{(F-H_0)/T}$ with $H_0$ as in
Eq.~(\ref{eq_Hamiltonian_0}). It can be generally assumed that the
latter's eigenvalues $E_{\mu}$, $\mu=1,2,3,4$, are non-degenerate
and the eigenstates are orthonormalized, $ \langle \nu |\mu
\rangle = \delta_{\mu\nu}$. Taking into account the qubits'
interaction with a dissipative
environment,~\cite{SmirnovRS,SmirnovDQ} we derive
\begin{equation}\label{eq_suscept}
 \chi_{zz}(\omega ) = - \sum_{\mu\neq \nu} \frac{\rho_{\mu} -\rho_{\nu}}
 {\omega + E_{\mu} - E_{\nu} + i\Gamma_{\mu\nu} }
P_{\mu \nu} ,
\end{equation}
\begin{equation}\label{eq_tan_theta}
\tan \chi = - 2\frac{Q_T}{L_T}  \sum_{\mu < \nu}\frac{ \rho_{\mu}
- \rho_{\nu}}{ E_{\nu} - E_{\mu} } P_{\mu \nu},
\end{equation}
where $\rho_{\mu} = \exp(-E_{\mu}/T)/Z$ is a thermal population of
the $\mu$ energy level, $Z = \sum_{\nu}\exp(-E_{\nu}/T),$
$\Gamma_{\mu\nu} $ are decoherence rates of the double-qubit
system,
 and
\begin{eqnarray}\label{eq_tan_theta_R}
P_{\mu \nu} &=& \lambda_a^2 \langle \mu | \sigma_{z}^{(a)}|\nu
\rangle \langle \nu |\sigma_{z}^{(a)}|\mu \rangle
+ \lambda_b^2 \langle \mu |\sigma_{z}^{(b)}|\nu \rangle
\langle \nu |\sigma_{z}^{(b)}|\mu \rangle\nonumber  \\
&&{}+ \lambda_a \lambda_b \langle \mu |\sigma_{z}^{(a)}|\nu
\rangle \langle \nu |\sigma_{z}^{(b)}|\mu \rangle\nonumber\\ &&{}+
\lambda_a \lambda_b \langle \mu |\sigma_{z}^{(b)}|\nu \rangle
\langle \nu |\sigma_{z}^{(a)}|\mu \rangle.
\end{eqnarray}
At low frequencies $\omega = \omega_T \ll |E_{\mu} - E_{\nu}|$ and
in a weak damping regime, $\Gamma_{\mu \nu}\ll |E_{\mu} -
E_{\nu}|$, the decoherence rates $\Gamma_{\mu \nu}$ have no effect
on $\tan\chi$, but are responsible for an equilibrium distribution in the system.

The first two terms in Eq.~(\ref{eq_tan_theta_R}) are non-zero
even if the two-qubit states are factorized. The first (second)
term corresponds to the contribution of qubit $a$ ($b$) and is
nonzero near the qubit's degeneracy point. These contributions
are practically independent of whether the qubits' degeneracy
points coincide or not.

The second and third lines in Eq.~(\ref{eq_tan_theta_R}) describe coherent
flipping of both qubits, which is \emph{only} possible for
non-factorizable (entangled) eigenstates $|\mu\rangle$,
$|\nu\rangle$. Therefore the difference between the coinciding IMT
dip of the two qubits and the sum of two single-qubit IMT dips
provides a measure of how coherent is the two-qubit dynamics (that
is, whether entangled eigenstates of the two-qubit Hamiltonian
Eq.~(\ref{eq_Hamiltonian_0}) are formed). This is a necessary
condition for the system to be in an entangled (pure or mixed)
state.

The measurement results are shown in Fig.~\ref{fig10}. Comparison
of the single-qubit dips with the coincident IMT dip shows clearly
that the contribution to $\tan \chi$ from the entangled
eigenstates is significant. Indeed, the amplitude of the central
dip in Fig.~\ref{fig10} at $T=50$~mK is 1.12, compared to the sum
of the single-qubit dips equal to 1.69. It means that the
entangled terms (second and third lines in Eq.~(\ref{eq_tan_theta_R})) are
responsible for the negative contribution $\approx-0.57$ to
$\tan\chi$.

At 50~mK the temperature is comparable to the characteristic
energies in the two-qubit system (at the two-qubit degeneracy
point the gap between the ground state and top excited state is
$\sim100$~mK). Since the characteristic measurement time in our
approach is dictated by the much smaller tank frequency,
$\omega_T$, the system will have time to equilibrate. Indeed, an
excellent quantitative agreement between the experiment
(Fig.~\ref{fig10}) and the theory Eq.~(\ref{eq_tan_theta})
confirms that the system is described by the equilibrium density
matrix with the Hamiltonian Eq.~(\ref{eq_Hamiltonian_0}) (all the
parameters of which we determined from the experiment). In other
words, our system is in an equilibrium mixture of entangled
two-qubit states.
\section{Summary}
We have shown that interferometer-type superconducting qubits can
be characterized by making use of the impedance measurement technique. Moreover, weak
continuous quantum measurements can be performed with this
method.
\section{Acknowledgments}
We are grateful to our colleagues M. H. S. Amin,  Ya. S.
Greenberg, H. E. Hoenig, U.~H\"ubner, A. Maassen van den Brink,
T.~May, V. I. Shnyrkov, and I. N.~Zhilyaev for their help and
contribution to this work on different stages.

Helpful discussions with D. V. Averin, G. Blatter, M. Feigel'man,
M. V. Fistul, V. B. Geshkenbein, A. J. Leggett, Yu. Makhlin, A.
N.~Omelyanchouk, A.~Shnirman, P. C. E. Stamp, S. Uchaikin, A. V.
Ustinov, A. D. Zaikin, and A. B. Zorin are gratefully
acknowledged.

\end{document}